\documentclass[aps,preprint,prd,nofootinbib]{revtex4}

\usepackage{amsmath,amssymb}
\usepackage{graphicx}
\usepackage{aas_macros}

\RequirePackage[colorlinks,citecolor=blue,urlcolor=blue]{hyperref}
\usepackage{soul}

\begin{document}

\title{Search for primordial black holes from gravitational wave populations using deep learning}

\author{Hai-Long Huang$^{1,2}$\footnote{\href{huanghailong18@mails.ucas.ac.cn}{huanghailong18@mails.ucas.ac.cn}}}
\author{Zhan-He Wang$^{2}$\footnote{\href{wangzhanhe19@mails.ucas.ac.cn}{wangzhanhe19@mails.ucas.ac.cn}}}
\author{Qing-Yu Lan$^{2}$\footnote{\href{lanqingyu19@mails.ucas.ac.cn}{lanqingyu19@mails.ucas.ac.cn}}}
\author{Jun-Qian Jiang$^{2}$\footnote{\href{jiangjunqian21@mails.ucas.ac.cn}{jiangjunqian21@mails.ucas.ac.cn}}}
\author{Jibin He$^{3}$\footnote{\href{20242701016@stu.cqu.edu.cn}{20242701016@stu.cqu.edu.cn}}}
\author{Yu-Tong Wang$^{1,2}$\footnote{\href{wangyutong@ucas.ac.cn}{wangyutong@ucas.ac.cn}}}
\author{Jun Zhang$^{4,2}$\footnote{Corresponding author:~\href{zhangjun@ucas.ac.cn}{zhangjun@ucas.ac.cn}}}
\author{Yun-Song Piao$^{1,2,4,5}$\footnote{Corresponding author:~\href{yspiao@ucas.ac.cn}{yspiao@ucas.ac.cn}}}

\affiliation{$^1$ School of Fundamental Physics and Mathematical
    Sciences, Hangzhou Institute for Advanced Study, UCAS, Hangzhou
    310024, China}

\affiliation{$^2$ School of Physical Sciences, University of
Chinese Academy of Sciences, Beijing 100049, China}

\affiliation{$^3$ Department of Physics and Chongqing Key Laboratory for Strongly Coupled Physics, Chongqing University, Chongqing 401331, P. R. China}

\affiliation{$^4$ International Center for Theoretical Physics
    Asia-Pacific, Beijing/Hangzhou, China}

\affiliation{$^5$ Institute of Theoretical Physics, Chinese
    Academy of Sciences, P.O. Box 2735, Beijing 100190, China}



\begin{abstract}

Gravitational waves (GWs) signals detected by the
LIGO/Virgo/KAGRA collaboration might be sourced (partly) by the
merges of primordial black holes (PBHs).
The conventional hierarchical Bayesian inference methods can
allow us to study population properties of GW events to search for
the hints for PBHs. However, hierarchical Bayesian analysis
require an analytic population model, and becomes increasingly
computationally expensive as the number of sources grows. In this
paper, we present a novel population analysis method based on deep
learning,
which enables the direct and efficient estimation of PBH
population hyperparameters, such as the PBH fraction in dark
matter, $f_{\rm PBH}$. Our approach leverages neural
posterior estimation combined with conditional normalizing flows
and two embedding networks. Our results demonstrate that
inference can be performed within seconds, highlighting the
promise of deep learning as a powerful tool for population
inference with an increasing number of GW signals for
next-generation detectors.

\end{abstract}

\maketitle

\tableofcontents

\section{Introduction}

First proposed by
Refs.\cite{Hawking:1971ei,Carr:1974nx,Zeldovich:1967lct},
primordial black holes (PBHs) have been a subject of extensive
research due to their potential implications for cosmological
evolution
\cite{Carr:2016drx,Chapline:1975ojl,Meszaros:1975ef,Carr:2020gox,Calza:2024fzo,Calza:2024xdh},
and the origin of supermassive black holes in galactic nuclei
\cite{Carr:2023tpt,Nakama:2016kfq}\footnote{Recent high-redshift
JWST observations seem to imply that some of those supermassive
black holes might be primordial,
e.g.\cite{Huang:2023chx,Huang:2024aog}, in particular the GHZ9 and
UHZ1 (at $z>10$) observed by the JWST, as well as Little Red Dots
(at $z>4$), can be explained naturally with supermassive
primordial black holes \cite{Hai-LongHuang:2024vvz,Hai-LongHuang:2024gtx}}. See also
e.g. Refs.
\cite{Garcia-Bellido:2017fdg,Sasaki_2018,annurev:/content/journals/10.1146/annurev-nucl-050520-125911,Carr:2021bzv,Green_2021,Escriva:2022duf,Domènech_2024,Khlopov:2008qy,Wang:2025hbw,Pi:2024ert,Jiang:2024aju,Zeng:2024snl,Lin:2022ygd,Lin:2021ubu,Wang:2024nmd,Cai:2023uhc,Zhu:2023gmx,Crescimbeni:2025ywm,Fakhry:2025mmt,Pritchard:2024vix,Stahl:2025mdu,Luo:2025ewp}
for further discussions. However, definitive evidence for the
existence of PBHs remains elusive. Gravitational wave (GW) signals
from PBH binaries could offer a direct probe of potential PBH
populations. These populations are expected to exhibit distinct
characteristics compared to astrophysical black hole (ABH)
populations, including differences in mass distribution
\cite{He:2023yvl,Andres-Carcasona:2024wqk}, the redshift
dependence of the merger rate
\cite{Stasenko:2024pzd,Raidal:2024bmm}, spin properties
\cite{DeLuca:2023bcr,DeLuca:2020bjf}, and even spatial
distribution
\cite{Huang:2023mwy,Crescimbeni:2025ywm,Clesse:2024epo}. Fully
utilizing the potential of GW observations is crucial for
enhancing the theoretical understanding of both populations (see
\cite{LISACosmologyWorkingGroup:2023njw} for recent review).


The conventional hierarchical Bayesian inference methods usually
are adopted when one used GW data to search for the evidence for
PBHs, which allow us to go beyond individual events to study
population properties
\cite{He:2023yvl,Huang:2024wse,Andres-Carcasona:2024wqk}. However,
hierarchical Bayesian analysis require an analytic population
model, and becomes increasingly computationally expensive as the
number of sources in the analysis grows, due both to the cost of
obtaining posterior samples for each event and the cost of
combining the events to derive the population posterior. Given the
large number of events expected from upcoming detector networks,
there is a growing need for new methods to efficiently measure PBH
population hyperparameters from GW events.

Thanks to the rapid advancements in machine learning, there has
been significant progress in the use and development of
simulation-based inference (SBI) approaches for data analysis
\cite{2020PNAS..11730055C,Brehmer:2020cvb,Bhardwaj:2023xph}. These
SBI methods differ in how they sample from the likelihood to
construct posterior densities or posterior samples, and are
classified into three categories: neural posterior estimation
(NPE) \cite{Wildberger:2022agw,Dax:2022pxd}, neural likelihood
estimation (NLE)
\cite{Alsing:2019xrx,Lin:2022ayr,2016arXiv160506376P}, and neural
ratio estimation (NRE)
\cite{Miller:2021hys,2021arXiv211000449R,2022arXiv220813624D,Miller:2022haf,2019arXiv190304057H,2020arXiv200203712D}.
They have been widely used in cosmology and astrophysics,
including CMB analysis \cite{Cole:2021gwr}, strong lensing image
analysis \cite{Montel:2022fhv}, point source searches
\cite{AnauMontel:2022ppb}, field-level cosmology
\cite{Makinen:2021nly}, and others
\cite{Dimitriou:2022cvc,Gagnon-Hartman:2023soa,Delaunoy:2020zcu,Karchev:2022xyn}.

In this work, we present a novel population analysis method based
on NPE that enables the direct and efficient estimation of PBH
population hyperparameters. We
model the population posterior distribution using a conditional
normalizing flow network \cite{tabak2010density, tabak2013family,
dinh2014normalizingflows, 2015arXiv150505770J,
KobyzevPAMI2020,2019arXiv191202762P}. There have been several
studies applying deep learning techniques to aspects of the
population inference problem
\cite{Talbot:2020oeu,Gerardi:2021gvk,Wong:2020jdt,Mould:2022ccw,Ruhe:2022ddi,Leyde:2023iof},
however, we are the first to apply this method to search for PBHs
from GW populations without the need for MCMC analysis. In
addition to the significant computational speed-up, our approach
avoids the explicit construction of complex likelihood functions,
only a realistic forward simulator to be provided. Moreover, it
provides an advantage in addressing potential biases that may
arise in the intermediate steps—biases that are challenging to
account for in conventional methods or when directly modeling the
population likelihood. For example, this includes the potential
variation in waveforms used for generating single-event posterior
samples \cite{Leyde:2023iof}.

The structure of this paper is organized as follows. In
Section~\ref{sec:BH}, we introduce the models used for both the
PBH and ABH binary populations, including their respective mass
functions and merger rates. Section~\ref{sec:HBA} provides a
review of the standard hierarchical Bayesian inference method,
which serves as a baseline for comparison. In
Section~\ref{sec:ML}, we describe the deep learning methodology
developed in this work in detail. Our main results are presented
in Section~\ref{sec:Results}. Finally, we conclude with a summary
and discussion of the broader implications in
Section~\ref{sec:conclusions}.

\section{Modelling BH binary populations}
\label{sec:BH}

\subsection{Primordial black hole population}

PBHs can form binaries through several formation channels both in the early Universe when two PBHs are produced sufficiently close to each other \cite{Bird:2016dcv,Sasaki:2016jop,Nakamura:1997sm}, and in the late Universe by capture in clusters
\cite{Bird:2016dcv,Nishikawa:2017chy,Ali-Haimoud:2017rtz},
as discussed in the recent review by \cite{Raidal:2024bmm}.
In this study, we focus on the formation channel of PBH binaries in the early
Universe, which is known to make a dominant contribution to the PBH merger
rate \cite{Raidal:2017mfl,Franciolini:2022ewd,Raidal:2024bmm}.
In this case,
the merger rate density per unit volume at cosmic time $t$ for PBHs is
\cite{Hai-LongHuang:2023atg,Liu:2018ess}
\begin{align}\label{eq:merger_rate_PBH}
    \frac{{\rm d}R_{\rm PBH}}{{\rm d}m_i{\rm d}m_j}&\approx\frac{1.99\times10^6}{\text{Gpc}^3\text{yr}}f^{1.46}
    \left(1+\frac{\sigma_{\text{eq}}^2}{f^2}\right)^{-0.27}
    \left(\frac{m_i}{M_\odot}\right)^{-0.92}\left(\frac{m_j}{M_\odot}\right)^{-0.92}
    \notag \\ & \times
    \left(\frac{m_i+m_j}{M_\odot}\right)^{0.97}
    \left(\frac{t}{t_0}\right)^{-0.92}\psi(m_i)\psi(m_j),
\end{align}
where $f\approx0.85f_{\rm PBH}$ is the total abundance of
PBHs in nonrelativistic matter, $t_0$ is the present time and
$\sigma_{\rm eq}^2$ is the variance of density perturbations
of the rest of dark matter at $z_{\rm eq}$.
We will focus only on PBHs in the stellar mass range, thus the effect of cosmic expansion on the comoving distance of PBH pairs is negligible \cite{Hai-LongHuang:2023atg}. In addition, we assume that PBHs are initially randomly distributed according to a spatial Poisson distribution. Generalizing to the case of initial clustering is straightforward \cite{Huang:2023mwy,Desjacques:2018wuu,Inman:2019wvr,DeLuca:2020jug}.

The PBH mass function in \eqref{eq:merger_rate_PBH} is defined by $\psi(m)\equiv\frac{m}{\rho_{\rm PBH}}\frac{{\rm d}n_{\rm PBH}}{{\rm d}m}$ normalized as $\int\psi(m){\rm d}m=1$. As an example, we consider a mass function where PBHs are sourced by supercritical bubbles that nucleated during slow-roll inflation
\cite{Huang:2023chx,Huang:2023mwy}
\begin{equation} \label{eq:bubbleMFa2}
    \psi_{\rm bubble}(m|M_c,\sigma)=e^{-\sigma^2/8}\sqrt{\frac{M_c}{2\pi\sigma^2 m^3}}\exp
    \left(-\frac{\ln^2(m/M_c)}{2\sigma^2}\right).
\end{equation}
As mentioned in the introduction, we use the deep learning tool \texttt{dingo} \footnote{We modified \texttt{dingo} to enable faster parallel inference: \url{https://github.com/JiangJQ2000/dingo}.} \cite{Dax:2021tsq}
to analyze the strain data and generate single-event posterior samples for the subsequent population analysis.

Since the current design of \texttt{dingo} does not guarantee that
faithfully its can be applied to all parameter ranges, especially
beyond those explored in its training set. we conservatively
consider only the events with (detector rest-frame) mass range
$m_1,m_2 \in [10,100]M_\odot$ and luminosity distance $d_L \in
[100,1000]$ Mpc. \footnote{In the following analysis, we focus
only on the distribution of redshift and mass, neglecting the spin
distribution. Therefore, the parameters defining each single GW
event are the masses of the PBH binary components, $m_1$ and
$m_2$, and the redshift $z$ (or equivalently, the luminosity
distance as we fixed our cosmology to Planck 2018
\cite{Planck:2018vyg}. Here, we do not consider the effect of the
Hubble tension on $H_0$ and relevant results (e.g.
\cite{Ye:2020btb,Ye:2021nej,Jiang:2021bab,Ye:2022efx,Jiang:2025hco,Jiang:2025ylr,Jiang:2024xnu,Jiang:2022uyg,Wang:2024dka}),
which may be actually negligible. }. This same selection is also
made for the case of the ABH model below. 
We anticipate that
future fast single-event strain methods will allow us to expand
their range.


\subsection{Astrophysical black hole population}

To describe the ABH population, one can use, for example, the wildly used
phenomenological POWER-LAW+PEAK model \cite{LIGOScientific:2018jsj,DeLuca:2021wjr,KAGRA:2021duu,Madau:2014bja} to model the differential merger rate of ABHs
${\rm d}R_{\rm ABH}/{{\rm d}m_i{\rm d}m_j}$
\begin{equation} \label{eq:merger_rate_ABH}
    \frac{{\rm d}R_{\rm ABH}}{{\rm d}m_1{\rm d}m_2}=R_{\rm ABH}^0(1+z)^\kappa
    p_{\rm ABH}^{m_1}(m_1)p_{\rm ABH}^{m_2}(m_2|m_1),
\end{equation}
where $R_{\rm ABH}^0$ is the local merger rate at redshift $z=0$, and $\kappa\simeq2.9$
describes the merger rate evolution with redshift \cite{KAGRA:2021duu,Madau:2014bja}.
The probability density function of the primary mass is modeled as a combination
of a power law and a Gaussian peak
\begin{equation}
    p_{\rm ABH}^{m_1}(m_1)=\left[(1-\lambda)P_{\rm ABH}(m_1)+\lambda G_
    {\rm ABH}(m_1)\right] S(m_1|\delta_m,m_{\rm min}),
\end{equation}
where
\begin{equation}
    P_{\rm ABH}(m_1|\alpha,m_{\rm min},m_{\rm max})\propto\Theta(m-m_{\rm min})
    \Theta(m_{\rm max}-m)m_1^{-\alpha},
\end{equation}
\begin{equation}
    G_{\rm ABH}(m_1|\mu_G,\sigma_G,m_{\rm min},m_{\rm max})\propto\Theta(m-m_{\rm min})
    \Theta(m_{\rm max}-m)\exp\left(-\frac{(m_1-\mu_G)^2}{2\sigma_G^2}\right)
\end{equation}
are restricted to masses between $m_{\rm min}$ and $m_{\rm max}$ and normalized.
The term $S(m_1|m_{\rm min},\delta_m)$ is a smoothing function, which rises from
0 to 1 over the interval $(m_{\rm min}, m_{\rm min}+\delta_m)$,
\begin{equation}
    S(m \mid m_{\rm min}, \delta_m) = \begin{cases}
        0, & m< m_{\rm min} \\
        \left[f(m - m_{\rm min}, \delta_m) + 1\right]^{-1}, & m_{\rm min} \leq m < m_{\rm min}+\delta_m \\
        1, & m\geq m_{\rm min} + \delta_m
    \end{cases}
\end{equation}
with
\begin{equation}
    f(m', \delta_m) = \exp \left(\frac{\delta_m}{m'} + \frac{\delta_m }{m' - \delta_m}\right).
\end{equation}
The distribution of the secondary mass is modelled as a power law,
\begin{equation}
    p_{\rm ABH}^{m_2}(m_2|m_1,\beta,m_{\rm min})\propto\left(\frac{m_2}{m_1}\right)^{\beta}.
\end{equation}
where the normalization ensures that the secondary mass is bounded by
$m_{\rm min}\le m_2\le m_1$.

In the later analysis, we shall consider two hypotheses: (1) All black hole binaries (BHBs) are of primordial origin; (2) All BHBs are of astrophysical origin. In the latter hypothesis,
the merger rate is given by \eqref{eq:merger_rate_ABH} while
in the former hypothesis, which we shall refer to as the PBH model, the merger
rate is \eqref{eq:merger_rate_PBH} with mass distribution given by \eqref{eq:bubbleMFa2}. In a more realistic scenario, BHBs could be either astrophysical or primordial, with the total merger rate given by ${\rm d}R/{\rm d}m_1{\rm d}m_2={\rm d}R_{\rm PBH}/{\rm d}m_1{\rm d}m_2+{\rm d}R_{\rm ABH}/{\rm d}m_1{\rm d}m_2$. We leave the exploration of this combined scenario, as well as other potential PBH and ABH models (e.g. \cite{PhysRevD.47.4244,Carr:2017edp,Yokoyama:1998xd,Niemeyer:1999ak,Musco:2012au,Carr:2016hva}), for future work.
The hyperparameters of our PBH/ABH model, along with the parameters describing each BHB, are listed in Table.~\ref{tab:1}.

\begin{table*}
     \centering
    \begin{ruledtabular}
        \begin{tabular} {cl}
            Event Parameter $\boldsymbol{\theta}$ & \\
            \hline \hline
            $m_1$ & Source-frame primary mass \\
            $m_2$ & Source-frame secondary mass \\
            $z$ & Merger redshift \\
          \hline \hline
            Hyperparameters $\Lambda$ & \\
            \hline \hline
            $M_c$ &  The characteristic mass \\
            $\sigma$ & The width of the distribution \\
            $f_{\rm PBH}$ & The fraction of dark matter in PBHs \\
            \hline
            $R_{\rm ABH}^0$ & Integrated merger rate of ABHs at $z=0$ \\
            $\lambda$ & Fraction of the Gaussian component in the primary mass distribution\\
            $m_{\rm min}$ & Minimum mass of the power low component in the primary mass
            distribution \\
            $m_{\rm max}$ & Maximum mass of the power low component in the primary mass
            distribution \\
            $\alpha$ & Inverse of the slope of the primary mass distribution for the power law component\\
            $\mu_G$ & Mean of the Gaussian component \\
            $\sigma_G$ & Width of the Gaussian component \\
            $\delta_m$ & Range of mass tapering on the lower end of the mass distribution\\
            $\beta$ & Spectral index for the power law of the mass ratio distribution
        \end{tabular}
    \end{ruledtabular}
    \caption{Event parameters $\boldsymbol{\theta}$ of the binary and hyperparameters $\Lambda$ of the PBH/ABH model used in this work. Detector-frame masses $m_d$ and source-frame masses $m_s$ are related as $m_d=(1+z)m_s$. We do not consider the mass growth of PBHs caused by accretion, nor the spin distribution.}\label{tab:1}
\end{table*}

\section{Hierarchical Bayesian population methods}
\label{sec:HBA}

The classical approach will function as our reference point against which
we will compare the outcomes with the deep learning methods.
Below, we refer to the classical method as HBA and to the deep learning model
as DL. In this section, we introduce the HBA methods, following the approach
outlined in \cite{He:2023yvl}.

According to the Bayes' theorem, the population posterior distribution
\begin{equation}
    p(\Lambda|\boldsymbol{d})=\frac{\mathcal{L}(\boldsymbol{d}|\Lambda)
    p(\Lambda)}{\mathcal{Z}_\Lambda},
\end{equation}
where we define the date measured from observed GW populations as $\boldsymbol{d}$. Hyper-prior $p(\Lambda)$ denotes the prior knowledge of $\Lambda$ (listed in Tab.~\ref{tab:2}) and
\begin{equation} \label{eq:evidence}
    \mathcal{Z}_{\Lambda}=\int\mathcal{L}(\boldsymbol{d}|\Lambda)
    p(\Lambda){\rm d}\Lambda
\end{equation}
is the hyper-evidence, the probability of
observing data $\boldsymbol{d}$.
The hyper-likelihood is
\cite{Loredo:2004nn,LIGOScientific:2020kqk,2019PASA...36...10T,Mandel:2018mve}
\begin{equation} \label{eq:tot_likelihood}
    \mathcal{L}(\boldsymbol{d}|\Lambda)\propto e^{-N(\Lambda)}\prod_{i=1}
    ^{N_{\rm obs}}\int T_{\rm obs}\mathcal{L}(d_i|\boldsymbol{\theta})\pi(\boldsymbol{\theta}|\Lambda){\rm d}\boldsymbol{\theta},
\end{equation}
Here,
$\mathcal{L}(d_i|\boldsymbol{\theta})$ is the single event likelihood, given some parameters
$\boldsymbol{\theta}$, and $\pi(\boldsymbol{\theta}|\Lambda)$ is called the hyperprior and governs the distribution of mass, spin, redshift and merger rate (the effect of the spin will be ignored in this work), given by
\begin{equation} \label{eq:population_model}
    \pi(\boldsymbol{\theta}|\Lambda)=\frac{1}{1+z}\frac{{\rm d}V_c}{{\rm d}z}
    \frac{{\rm d}R}{{\rm d}m_1{\rm d}m_2}(\boldsymbol{\theta}|\Lambda),
\end{equation}
where ${\rm d}V_c/{\rm d}z$ represents the
differential comoving volume, and $\boldsymbol{\theta}\equiv\{m_1,m_2,z\}$ constitutes the
parameters that defining the GW event.
$N(\Lambda)=T_{\rm obs}\xi(\Lambda)$ represents the Poisson probability of observing the expected number of detections over the observation timespan $T_{\rm obs}$,
with the selection bias $\xi(\Lambda)$ accounting for the selection biases introduced by the detector’s sensitivity
\begin{equation}
    \xi(\Lambda)=\int p_{\rm det}(\boldsymbol{\theta})\pi(\boldsymbol{\theta}|\Lambda){\rm d}\boldsymbol{\theta},
\end{equation}
where $p_{\rm det}(\boldsymbol{\theta})$ denotes the detection probability,
and depends primarily on the masses and redshift of the system \cite{LIGOScientific:2020kqk}.
The estimation of $\xi(\Lambda)$ is performed using simulated
injections samples, where
a Monte Carlo integral over $N_{\rm inj}$ injections samples is used to
approximate it as
\begin{equation}
    \xi(\Lambda)\approx\frac{1}{N_{\rm inj}}\sum_{j=1}^{N_{\rm det}}
    \frac{\pi(\boldsymbol{\theta}_j|\Lambda)}{\pi_{\rm inj}(\boldsymbol{\theta}_j)}\equiv\frac{1}
    {N_{\rm inj}}\sum_{j=1}^{N_{\rm det}}s_j,
\end{equation}
where $\pi_{\rm inj}(\boldsymbol{\theta}_j)$ is the prior probability of the $j$th event
(the probability density function from which the injections are drawn),
$N_{\rm det}$ denoting the count of successfully detected injections samples.



\begin{table*}
     \centering
        \begin{tabular} {cccccc}
            \hline \hline
            Parameter & Unit & Prior & True value & Posterior(HBA) & Posterior(DL) \\
            \hline \hline
            $M_c$ & $M_\odot$ & U$(20,70)$ &40.97& $40.42^{+0.87}_{-0.80}$ & $40.53^{+0.98}_{-0.98}$ \\
            $\log \sigma$ & - & U$(-1,0)$ &-0.70& $-0.74^{+0.05}_{-0.05}$ & $-0.72^{+0.07}_{-0.07}$ \\
            $\log f_{\rm PBH}$ & - & U$(-5,0)$ &-3.01& $-3.08^{+0.03}_{-0.03}$ & $-3.02^{+0.01}_{-0.01}$\\
            \hline
            \hline
            $\log R_{\rm ABH}^0$ & ${\rm Gpc}^{-3}{\rm yr}^{-1}$ & U$(-2,3)$ &-1.13& $0.065_{-0.62}^{+1.85}$ & $-2.07_{-0.51}^{+0.53}$ \\
            $\log \lambda$ & - & U$(-6,0)$ &-5.31& $-3.76_{-2.02}^{+3.76}$ & $-3.97_{-1.38}^{+1.88}$\\
            $m_{\rm min}$ & $M_\odot$ & U$(10,20)$ &10.96& $10.53_{-0.40}^{+0.80}$ &$10.97_{-0.61}^{+1.43}$ \\
            $m_{\rm max}$ & $M_\odot$ & U$(40,100)$ &86.69& $69.32_{-20.97}^{+20.52}$ & $69.53_{-21.71}^{+23.01}$\\
            $\alpha$ & - & U$(-4,12)$ &10.94& $11.82_{-0.67}^{+0.14}$ & $11.50_{-0.94}^{+0.81}$\\
            $\mu_G$ & $M_\odot$ & U$(20,50)$ &26.29& $21.00_{-0.98}^{+12.38}$ & $33.09_{-8.93}^{+11.24}$ \\
            $\sigma_G$ & $M_\odot$ & U$(1,10)$ &2.62& $1.57_{-0.56}^{+5.27}$& $3.98_{-2.54}^{+3.92}$\\
            $\delta_m$ & $M_\odot$ & U$(0,10)$ &3.39& $3.92_{-2.67}^{+2.99}$ & $5.33_{-4.10}^{+3.20}$ \\
            $\beta$ & - & U$(-4,12)$ &-1.99& $-3.87_{-0.09}^{+0.44}$ & $2.52_{-4.87}^{+6.05}$\\
            \hline \hline
        \end{tabular}
    \caption{Prior and posterior credible intervals $(84\%)$ of both our PBH model and the POWER LAW+PEAK ABH model are shown. We simulate a detected population of 64 events, with posterior samples for each event obtained using \texttt{dingo}. These posterior samples, together with the observation time computed via Eq.~\eqref{eq:t_obs}, serve as the input to both the HBA and ML (deep learning) methods for hyperparameter inference.}\label{tab:2}
\end{table*}

On the other hand, the single event likelihood is usually not available.
Instead, posterior samples are provided.
The intergral appearing in \eqref{eq:tot_likelihood} for each
event can be approximated using Monte-Carlo integration as \cite{LIGOScientific:2020kqk}
\begin{equation}
    \int\mathcal{L}(d_i|\boldsymbol{\theta})\pi(\boldsymbol{\theta}|\Lambda){\rm d}\boldsymbol{\theta}\approx
    \frac{1}{n_i}\sum_{j=1}^{n_i}\frac{\pi(\boldsymbol{\theta}_{ij}|\Lambda)}
    {\pi_{\varnothing}(\boldsymbol{\theta}_{ij}|\Lambda)}\equiv\frac{1}{n_i}
    \sum_{j=1}^{n_i}\omega_{ij},
\end{equation}
where $\boldsymbol{\theta}_{ij}$ denotes the intrinsic parameters of the
$j$th sample ($n_i$ posterior samples totally) of the $i$th event,
and $\pi_{\varnothing}(\boldsymbol{\theta}_{ij}|\Lambda)$ is the prior used for the initial
parameter estimation. The sum above is taken over the posterior samples
$\boldsymbol{\theta}_{ij}\sim p(\boldsymbol{\theta}_{ij}|d_i)$.

Finally, the log-likelihood is evaluated as \cite{Mastrogiovanni:2023zbw}
\begin{equation}
    \ln \mathcal{L}(\boldsymbol{d}|\Lambda)\approx-\frac{T_{\rm obs}}{N_{\rm inj}}
    \sum_{j}^{N_{\rm det}}s_j + \sum_i^{N_{\rm obs}}\ln\left(\frac{T_{\rm obs}}{n_i}
    \sum_j^{n_i}\omega_{ij}\right).
\end{equation}
The posterior of the hyperparameters $\Lambda$ given the observed dataset $\boldsymbol{d}$ is obtained by \texttt{emcee} \cite{Foreman-Mackey:2012any}.

Model selection is a common challenge in GW population analyses. To distinguish between competing models, such as PBH and ABH models, the Bayes factor is often used. The log-Bayes factor comparing two models is defined as
\begin{equation}
    \ln \mathcal{B}^{\rm PBH}_{\rm ABH}=\ln\mathcal{Z}_{\rm PBH} - \ln\mathcal{Z}_{\rm ABH},
\end{equation}
where the Bayesian evidence (or marginal likelihoods) for a model $\mathcal{Z}_{\rm \Lambda}$ is computed as \eqref{eq:evidence}. The sign of log-Bayes tells us which model is preferred.
When the absolute value of it is large, we say that one model is preferred over the other.
In conventional approaches, the computation of Bayesian evidence requires resource-intensive techniques such as Markov Chain Monte Carlo (e.g., \texttt{emcee}) or nested sampling (e.g., \texttt{dynesty} \cite{Speagle:2019ivv}), both of which can become computationally expensive, particularly when dealing with high-dimensional parameter spaces or large datasets.

\section{Deep learning methods}
\label{sec:ML}

\begin{figure}[htbp]
    \centering
    \includegraphics[width=1\textwidth]{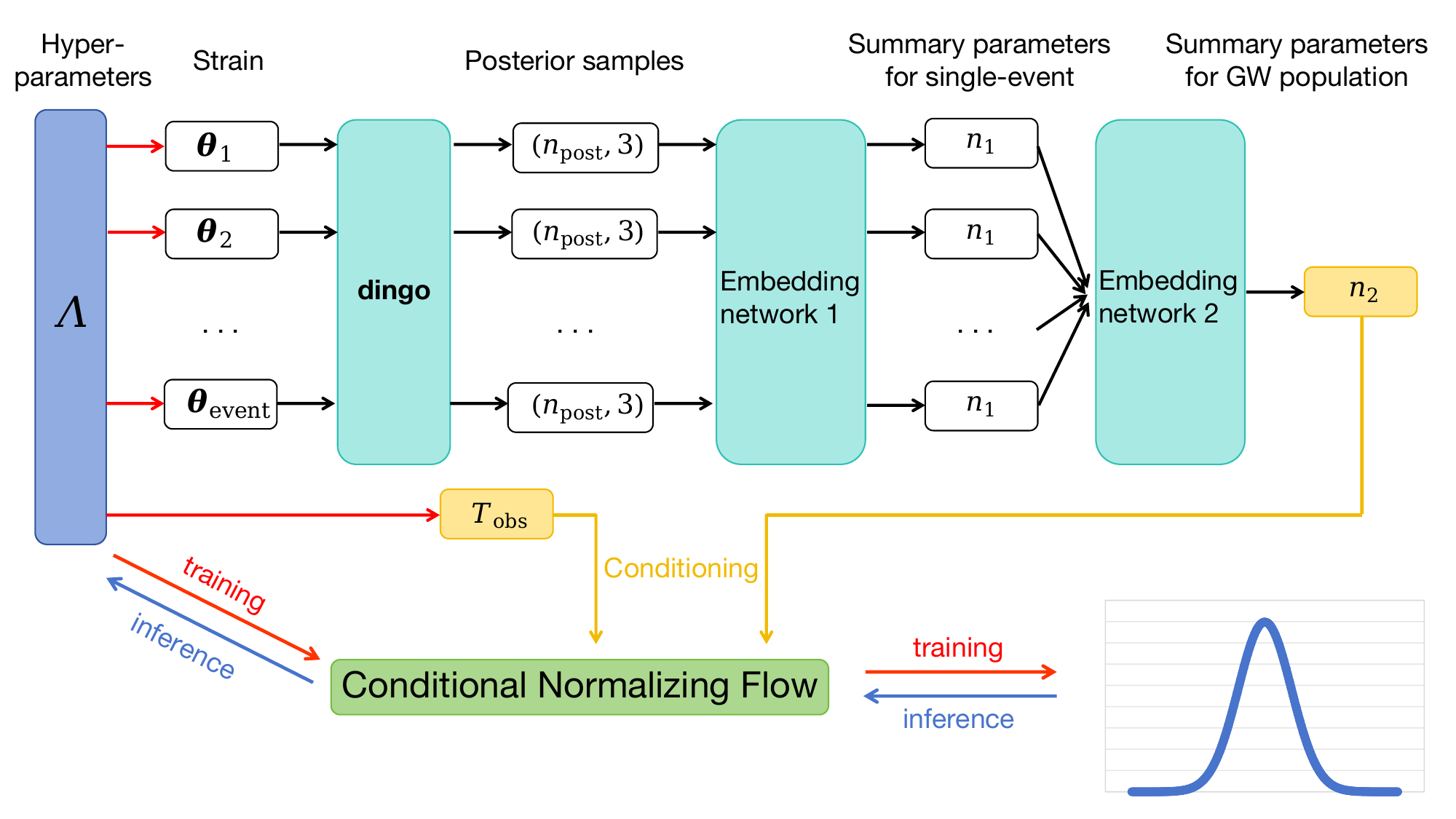}
    \caption{Schematic overview of the deep learning model which is composed of a normalizing flow and two embedding neural networks. Solid arrows represent input-output relations: red apply during training, blue ones when performing inference while black and orange ones are always present.}
    \label{fig:CNF}
\end{figure}

In this section, we describe how we apply the NPE method to directly estimate the posterior density, enabling fast and accurate hierarchical Bayesian inference for gravitational wave populations.

\subsection{Neural network architecture}
\label{subsec:FA}

\begin{table*}
     \centering
        \fontsize{9}{0}
        \begin{tabular} {ccc}
            \hline \hline
            Model & PBH & ABH  \\
            \hline \hline
            Dimensions of hyperparameters of the physical model & 3 & 9 \\
            Events per batch $n_{\rm event}$ & 64 & 64 \\
            Posterior samples per event $n_{\rm post}$ & 128 & 128 \\
            Dimensions embedding network 1 & $(512, 256, 256, 128, 128, 64)$ & $(512, 256, 256, 256, 128, 128, 64)$ \\
            Dimensions embedding network 2 & $(256, 256, 128, 128, 64)$ & $(512, 512, 256, 256, 128)$ \\
            Number of residual blocks in the NF & 5 & 13 \\
            Number of hidden units in the NF & 24 & 104 \\
            Training epochs & 200 & 200 \\
            Initial learning rate & 0.001 & 0.001 \\
            Batch size & 256 & 256 \\
            Training population samples & $5 \times 10^3$ & $1 \times 10^4$ \\
            \hline \hline
        \end{tabular}
    \caption{Architecture of the embedding networks and the NF for PBH and ABH model.}\label{tab:NFparameter}
\end{table*}

Our method uses a normalizing flow (NF) combined with two embedding neural networks for data compression to directly estimate population hyperparameters from a collection of individual source observations. The general structure of our pipeline is depicted in Fig.~\ref{fig:CNF}, showing the input/output relations between its building blocks.

The pipeline takes as input $n_{\rm event} = 64$ events \footnote{Here, we fix the number of events in a single neural network (NN) analysis, since NNs typically require a fixed input size. To handle more events, we can divide them into sub-groups, as will be discussed later.},
each described by whitened strain time series and their observation time $T_{\rm obs}$, and $n_{\rm event} \times n_{\rm post}$ outputs samples ($n_{\rm post} = 128$ samples per event) from the posterior probability distribution of the component masses and luminosity distance generated from \texttt{dingo} \footnote{However, even if the \texttt{dingo} algorithm is not a perfect approximation to the single event posterior, this does not invalidate our approach. By construction, the DL model learns the posterior distribution marginalized over the \texttt{dingo} uncertainty. While it is possible to correct for potential inaccuracies in \texttt{dingo} using importance sampling, this would significantly increase the computation time in the process of training our networks. In addition to the single event posterior samples, the method could be applied to any input data that summarizes the GW observations sufficiently well, e.g. time series data and spectrogram visualizations.}.
We use posterior samples that have been standardized by subtracting their mean and dividing by their standard deviation for each variable, which helps accelerate model convergence.
Embedding network 1 compresses each event (standardized posterior samples) into $n_1$ summary parameters.
These $n_{\rm event} \times n_1$ scalars are then further compressed by Embedding network 2 into $n_2$ scalars. Together with $T_{\rm obs}$, they are fed into the flow model.
The use of the two embedding networks significantly reduces the number of free parameters in the model, helping to mitigate overfitting. Specificially, the architecture of the embedding networks is composed of several fully connected layers, each followed by a ReLU activation function to introduce non-linearity.

For PBH, embedding network 1 has 6 hidden layers with (512, 256, 256, 128, 128, 64) neural units for each hidden layer and embedding network 2 has 5 hidden layers with (256, 256, 128, 128, 64) neural units for each hidden layer.
While for ABH model, which is a much more complex model, embedding network 1 has 7 hidden layers with (512, 256, 256, 256, 128, 128, 64) neural units for each hidden layer and embedding network 2 has 5 hidden layers with (512, 512, 256, 256, 128) neural units for each hidden layer.
Furthermore, to prevent over-fitting in the case of ABH, we additionally added dropout layers with a probability of 0.4 between each layer of the embedding network 2.
We summarize the
details of the specific network in Tab.~\ref{tab:NFparameter}.

The core of the model is the normalizing flow (NF), which is responsible for reconstructing the posterior distribution. A brief introduction to the NF is
provided below.

NFs offer an effective method for representing complex probability distributions using neural networks. This approach facilitates efficient sampling and density estimation by expressing the distribution as a sequence of mappings, or "flows", $f : u \to \Lambda$, which maps from a simpler base distribution $u$ (typically a standard normal distribution $\mathcal{N}(0,1)$) to the parameter space, which in our case corresponds to the hyperparameters of the PBH/ABH model. When the mapping $f$ depends on the observed data, denoted as $f_{\boldsymbol{d}}$, it describes a conditional probability distribution $q(\Lambda | \boldsymbol{d})$. The probability density function is given by the change of variables formula:
\begin{equation} \label{eq:A1}
    q(\Lambda|\boldsymbol{d})=\mathcal{N}(0,1)^D(f_{\boldsymbol{d}}^{-1}(\Lambda))|{\rm det}f_{\boldsymbol{d}}^{-1}(\Lambda)|,
\end{equation}
where $D$ is the dimensionality of the parameter space. The term “normalizing flows” refers to the sequence of transformations that progressively ``normalize" the distribution into the desired complex form.

It is essential to design and combine multiple transformations to model complex distribution. Each transformation should be both invertible (so that $f_{\boldsymbol{d}}^{-1}(\Lambda)$ can be evaluated for any $\Lambda$) and possess a tractable Jacobian determinant (allowing for efficient computation of ${\rm det}f_{\boldsymbol{d}}^{-1}(\Lambda)$). These properties enable efficient sampling and density estimation, as described in equation \eqref{eq:A1}. Various normalizing flow architectures have been developed to satisfy these conditions, typically by composing several simpler transformations $f^{(j)}$, with each transformation being parameterized by the output of a neural network. To sample from the posterior, $\Lambda \sim q(\Lambda | \boldsymbol{d})$, we first sample $u \sim \mathcal{N}(0,1)^D$ and then apply the flow in the forward direction, as illustrated by the blue arrows in Fig.~\ref{fig:CNF}.

For each flow step, we employ a conditional coupling transformation
\cite{2019arXiv190604032D}. In this setup, the first $k$ components of the input are fixed, while the other undergo an elementwise transformation conditioned on all components and the data:
\begin{equation}
    f_{\boldsymbol{d},i}^{(j)}(u)=
    \begin{cases}
        u_i & \text{if } i\le k,
        \\
        f_i^{(j)}(u_{1:D},\boldsymbol{d}) &  \text{if } k< i\le D.
    \end{cases}
\end{equation}
When the elementwise functions $f_i^{(j)}$ are chosen to be monotonic, quadratic, rational spline functions, the transformation inherently satisfies the conditions required for a normalizing flow. The spline parameters for each $f_i^{(j)}$ are learned from the neural network's output, which takes as input the concatenated data, $u_{1:D}$ and $\boldsymbol{d}$. To maintain the flexibility of the entire flow, we randomly permute the parameters between each transformation.
Before the vectors obtained by embedding network 2 are input into this spline function, they will also pass through a residual network.
The detail of the NF is summarized in~\autoref{tab:NFparameter}.
We utilize the implementation of this structure provided by \texttt{normflows} \footnote{\url{https://github.com/VincentStimper/normalizing-flows}.} \cite{Stimper2023}.

\subsection{Training the networks}

Fitting a flow-based model $q(\Lambda | \boldsymbol{d})$ to a target distribution $p(\Lambda | \boldsymbol{d})$
can be done by minimizing some divergence or discrepancy between them.
One of the most popular choices is the forward Kullback-Leibler (KL) divergence that is mass-covering \cite{Kullback:1951zyt},
\begin{equation}
    D_{\rm KL}(p||q)=\int{\rm d}\Lambda p(\Lambda|\boldsymbol{d})\log\frac{p(\Lambda|\boldsymbol{d})}{q(\Lambda|\boldsymbol{d})}.
\end{equation}
This measure indicates how much information is lost when using $q$ as an approximation to $p$. The forward KL divergence is well-suited for situations in which we have samples
from the target distribution (or the ability to generate them), but we cannot necessary
evaluate the target density. By taking the expectation over data samples $\boldsymbol{d} \sim p(\boldsymbol{d})$, we can simplify the expression, resulting in the loss function:
\begin{align}
    L&= \int {\rm d}\boldsymbol{d}p(\boldsymbol{d})\int{\rm d}\Lambda p(\Lambda|\boldsymbol{d})\log\left(\frac{p(\Lambda|\boldsymbol{d})}{q(\Lambda|\boldsymbol{d})}\right) \notag \\
    &=\int {\rm d}\boldsymbol{d}p(\boldsymbol{d})\int{\rm d}\Lambda p(\Lambda|\boldsymbol{d})[-\log q(\Lambda|\boldsymbol{d})]+\text{constant.} \notag \\
    &=\int{\rm d}\Lambda p(\Lambda)\int {\rm d}\boldsymbol{d}p(\boldsymbol{d}|\Lambda)[-\log q(\Lambda|\boldsymbol{d})]+\text{constant.}
\end{align}
On the third line, we applied Bayes’ theorem to rewrite the cross-entropy between the two distributions in terms of the likelihood, thereby avoiding dependence on the unknown true posterior. Finally, the loss function can be approximated on a mini-batch of samples, omitting constant terms that are independent of the flow's parameters,
\begin{equation}
    L\approx-\frac{1}{N}\sum_{i=1}^{N}\log q(\Lambda^{(i)}|\boldsymbol{d}^{(i)}),
\end{equation}
where $N$ samples are drawn ancestrally in a two-step process: sample from the prior, $\Lambda^{(i)}\sim p(\Lambda)$ and simulate data according to the population model, $\boldsymbol{d}^{(i)}\sim p(\boldsymbol{d}|\Lambda^{(i)})$.
Minimizing the above Monte Carlo approximation of the loss function is equivalent to
fitting the flow-based model to the samples by maximum likelihood estimation.

In order to avoid over-fitting, we add a $L^2$ regularization term to the loss function:
\begin{equation}
    L = -\frac{1}{N}\sum_{i=1}^{N}\log q(\Lambda^{(i)}|\boldsymbol{d}^{(i)}) + \lambda \sum_\text{NN}{w},
\end{equation}
where $\sum_\text{NN}{w}$ is the sum of weights in the network and $\lambda$ is choosen  to be $1 \times 10^{-5}$ for the PBH model.
For the ABH model, we set $\lambda = 1 \times 10^{-3}$ as it is likely to be over-fitted for small training sets.
We then take the gradient of $L$ with respect to network parameters and minimize using the Adam optimizer \cite{2014arXiv1412.6980K}.
The learning rate is started from $10^{-3}$ and is reduced during training using the Plateau scheduler.

In our work,
the parameters of two embedding networks implicitly appear in the loss function
and are thus optimized jointly with the parameters defining the flow transformation.
The batch size is 256 for each model.

\subsection{Generating the training set}

For computational reasons, we precompute the samples $\Lambda_i$ from the prior
$p(\Lambda)$ and for each sample $\Lambda$, we
draw 128 events, characterized by $\{m_1, m_2, d_L\}$ for each event, according to the PBH/ABH population model \eqref{eq:population_model} for each sample $\Lambda$.
We simulate thier
observed strains (passing some specified selection threshold) and produce
256 posterior samples for each sample $\Lambda$ with \texttt{dingo}.
Specifically, we sample a event according to the population model $\pi(\boldsymbol{\theta}|\Lambda)$, generate its waveform, and compute its signal-to-noise ratio (SNR) \footnote{Here, we assume the O1 sensitivity curve \cite{LIGOScientific:2014pky} for the Laser Interferometer Gravitational Wave Observatory (LIGO) Hanford and LIGO Livingston detectors.
As noted in \cite{Leyde:2023iof}, this is an approximation, as the application of this method to real data will require more intricate selection criteria, such as incorporating the false alarm rate. To fully account for selection effects, an injection campaign would be necessary, similar to the one conducted in the HBA methods discussed in Section~\ref{sec:HBA}.}.
This process continues until we obtain 256 events with SNR $> {\rm SNR}_c = 12$. In this process, the total number of samples generated is $N(\Lambda)$.
Therefore, the required observation time $T_{\rm obs, 128}$ for the 128 events, which is crucial for inferring the the average number density of BHBs in the PBH/ABH population model later, is given by
\begin{equation} \label{eq:t_obs}
    T_{\rm obs,128}(\Lambda)=\frac{N(\Lambda)}{\int\pi(\boldsymbol{\theta}|\Lambda){\rm d}\boldsymbol{\theta}}.
\end{equation}
By construction, the model contains the selection
effect term $\xi(\Lambda)$. We thus avoid the computation of this term during
inference.

During each training epoch, for each hyperparameter $\Lambda$, we randomly take $n_\text{event} = 64$ out of the 128 events we generated.
Meanwhile, for each event, we also randomly take $n_\text{post} = 128$ out of the 256 samples we generated.
These can help prevent overfitting.
Combined with the observation time for $n_\text{event} = 64$, $T_{\rm obs} = T_{\rm obs,128} / 2$, Our training dateset contains $(n_{\rm event} \times n_{\rm post} \times 3) + 1$ scalars for each $\Lambda$.
Due to the limitation of computational resources, we generated $4 \times 10^3$ population samples (i.e., $\Lambda$) for the PBH model and $1 \times 10^4$ population samples for the ABH model.

\subsection{Split sub-populations}

A major complication is that the number of events to be observed is typically not known a prior and will increase over time while NNs typically require a fixed input dimension. This issue can be addressed using the ``divide-and-conquer" strategy proposed by Ref. \cite{Leyde:2023iof}, which splits the population into smaller sub-populations for independent analysis, and then merges them to obtain the final result by analyzing the complete catalog through importance sampling. Specifically, the GW population is devided into smaller sub-populations of events, $\boldsymbol{d} = \bigcup_{i=1}^{n_b} d_{K_i}$, where each $d_{K_i}\equiv\{d_k\}_{k\in K_i}$ contains $n_{\rm event}$ events, such that $n_b \equiv N_{\rm obs}/n_{\rm event}$. Here, we assume the number of events in each subset, $n_{\rm event}$, divides the total number of observed events $N_{\rm obs}$.
The deep learning model generates a population posterior, $q(\Lambda|d_{K_i})$, for each sub-population, which approximates $p(\Lambda|d_{K_i})$. The complete posterior is obtained by combining the individual posteriors of each sub-population (see Ref.~\cite{Leyde:2023iof} for further details). This approach also ensures that the computational cost of generating the training dataset remains manageable.



\begin{figure}[htbp]
    \centering
    \includegraphics[width=0.48\textwidth]{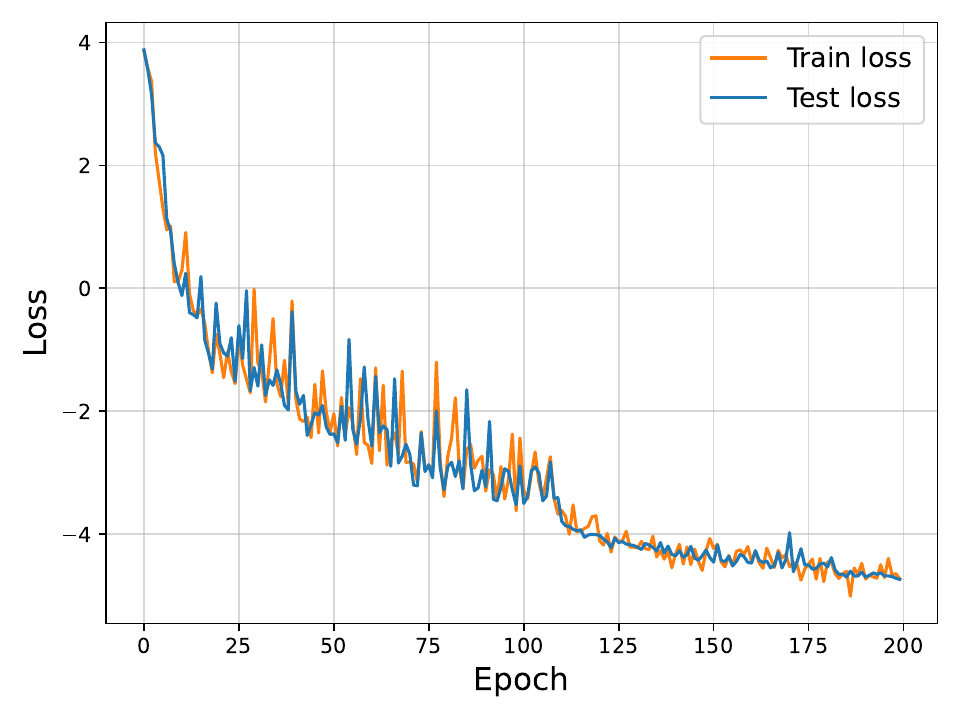}
    \quad
    \includegraphics[width=0.48\textwidth]{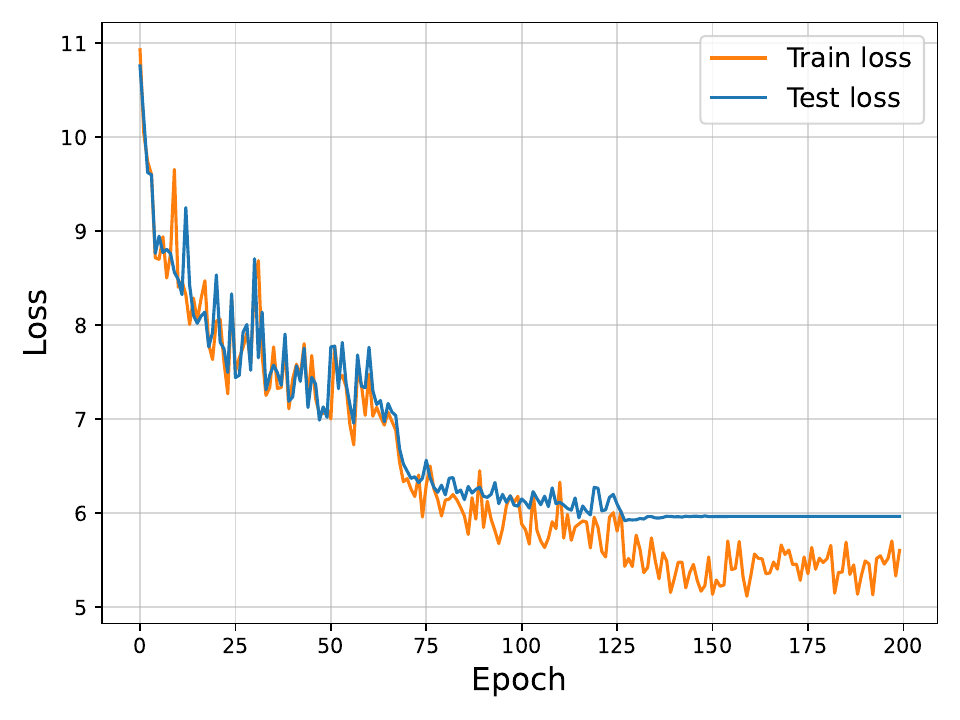}
    \caption{Loss for the PBH (left) and ABH (right) model. Since the test loss (blue) and train loss (orange)
    do not differ much, we conclude that the model can generalize effectively to data that were not included in the optimization process.}
    \label{fig:loss_curve}
\end{figure}

\section{Results}
\label{sec:Results}


The training time of the model was only 30 minutes on an NVIDIA V100S GPU.
The associated
training and test loss curves of model are plotted in Fig.~\ref{fig:loss_curve}.
The train and test loss coincide, suggesting tht the model can process unseen input data and generate accurate hyperparameter posterior distributions.

\begin{figure}[htbp]
    \centering
    \includegraphics[width=0.48\textwidth]{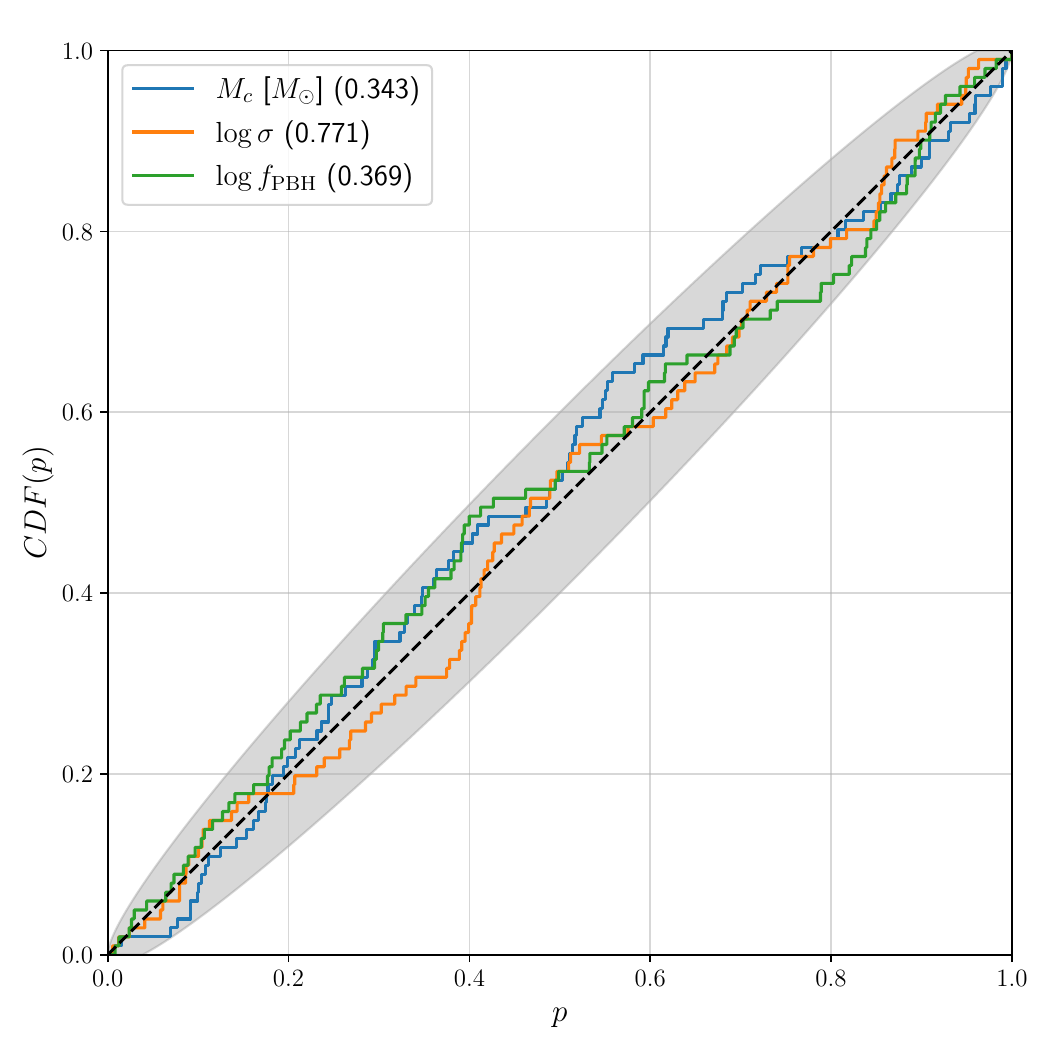}
    \quad
    \includegraphics[width=0.48\textwidth]{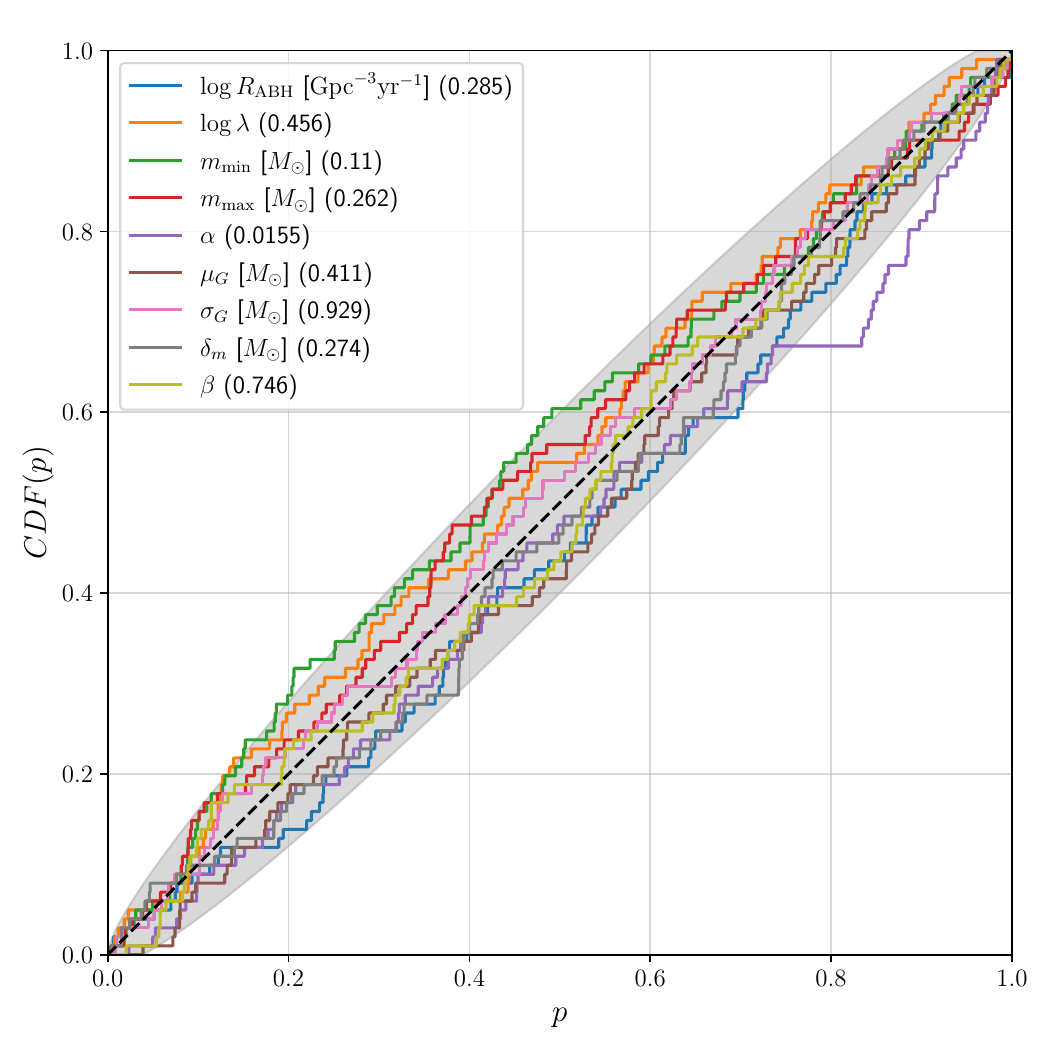}
    \caption{Probability-Probability plot for a set of 1024 posterior evaluations from the training set for the PBH (left) and ABH (right) model. Each cumulative distribution aligns well with the diagonal, with the spread mostly confined within the $2\sigma$ gray regions for almost the entire confidence level interval. The legend shows the $p$-values of the individual parameters, with a combined $p$-value of 0.59 (0.14) for PBH (ABH) model, implying that the network
    has correctly learned the desired posterior distribution.}
    \label{fig:pp_plot}
\end{figure}

As a test, we evaluate our model on data that is fully consistent with the training distribution. We draw posterior samples from 1024 simulated datasets and construct a P-P plot shown in Fig.~\ref{fig:pp_plot}. For each hyperparameter, we compute the percentile score of the true value within its marginalized posterior, and plot the cumulative distribution function (CDF) of these scores. For the true posteriors, the percentiles should be uniformly distributed, meaning the CDF should follow a diagonal line. The Kolmogorov-Smirnov test $p$-values are reported in the legend, ranging from $35.2\%$ to $92.6\%$ ($1.55\%$ to $92.9\%$) for the PBH (ABH) model, with a combined $p$-value of 0.59 (0.14). This indicates that our model correctly reconstructs the population posterior.

\begin{figure}[htbp]
    \centering
    \includegraphics[width=0.8\textwidth]{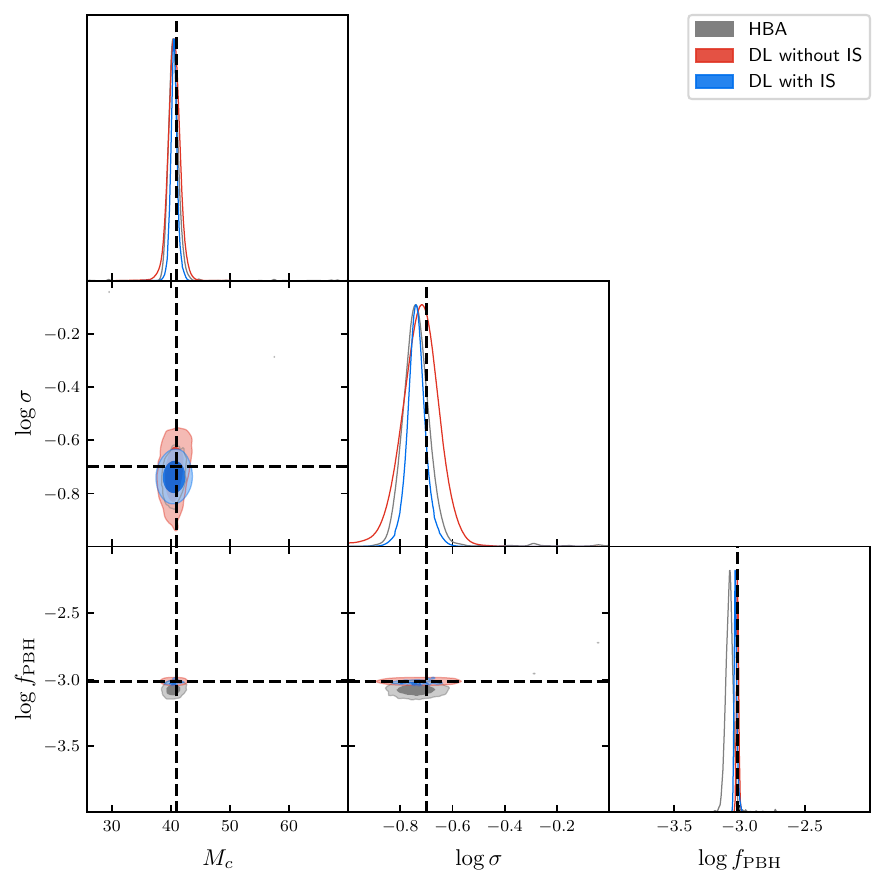}
    \caption{Results from deep learning method (DL without IS, red) compared to a conventional hierarchical Bayesian analysis (HBA, grey) for PBH model. The posterior is inferred from a GW population consisting of $n_{\rm event}=$64 events, with both analyses using \texttt{dingo} samples as input data. The deep learning method outperforms the traditional Bayesian approach in estimating $\log f_{\rm PBH}$, as it produces a narrower distribution (i.e., smaller variance) while remaining centered around the injected value. In other cases, discrepancies with the conventional HBA results can be reduced by applying importance sampling with the classical likelihood. The reweighted posterior (DL with IS, blue) closely matches the classical result for $M_c$ and $\log \sigma$.}
    \label{fig:infer}
\end{figure}

\begin{figure}[htbp]
    \centering
    \includegraphics[width=1\textwidth]{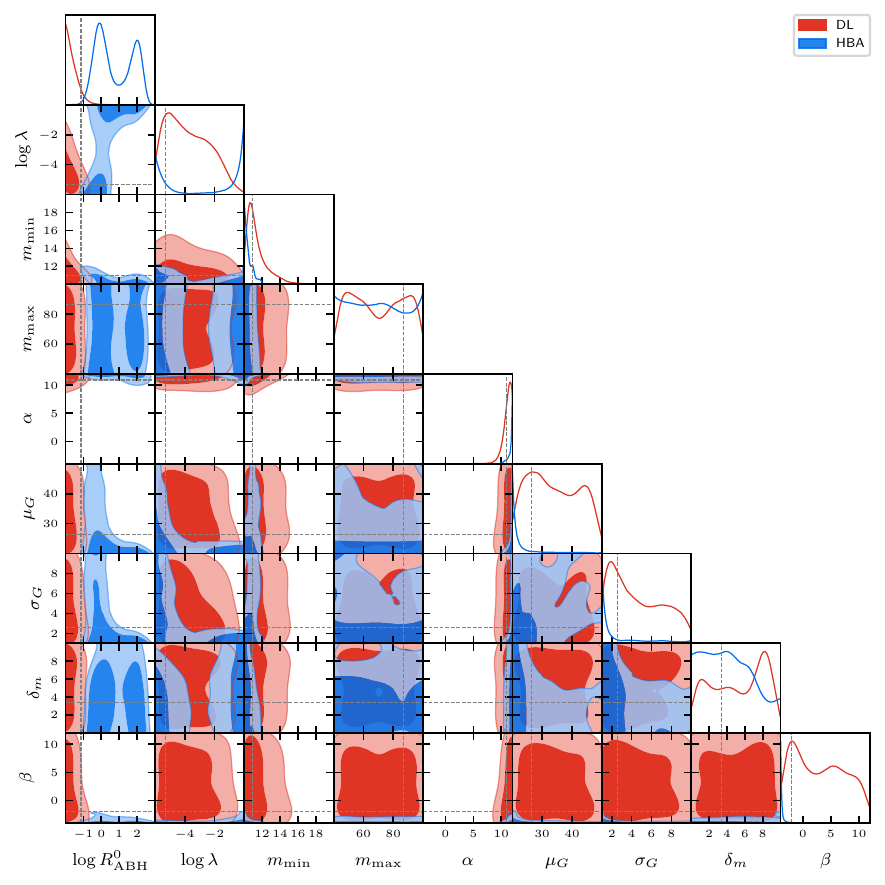}
    \caption{Results from deep learning method (DL, red) compared to a conventional hierarchical Bayesian analysis (HBA, blue) for ABH model.}
    \label{fig:infer_ABH}
\end{figure}

We now proceed to our main result, which is a comparison of the deep learning method with traditional HBA methods. Fig.~\ref{fig:infer} and Fig is representative of the majority of cases, where we generally observe good agreement between the two methods (The one- and two-sigma intervals are summarized in Tab.~\ref{tab:2}.). The deep learning approach allows us to generate an effective sample size of $2^{14}$ posterior samples in just 5 seconds of computation time. In contrast, it takes more than ten minutes to generate the same sample points using the conventional HBA method, even with parallel computation on multiple nodes. This highlights the significant computational advantage of the deep learning approach over traditional methods. Moreover, we find that the deep learning method outperforms the traditional HBA approach in some cases. For example, it can be seen from Fig.~\ref{fig:infer} that the deep learning method is particularly effective in estimating $\log f_{\rm PBH}$.
We attribute this to the fact that $f_{\rm PBH}$ represents the average number density of PBHs, and this set of properties is almost perfectly captured by just one parameter: the observation time $T_{\rm obs}$ for $n_{\rm event}$ enents. This parameter is well-learned by the neural network, allowing it to efficiently capture the underlying distribution. As a result, the neural network exhibits exceptional performance in inferring $f_{\rm PBH}$, leading to superior behavior in comparison to traditional methods. In other cases, the discrepancies between the deep learning results and the HBA approach can be addressed by reweighting the NPE samples to match the target HBA posterior using importance sampling weights:
\begin{equation}
    \omega(\Lambda)\propto\frac{\mathcal{L}(\Lambda|\boldsymbol{d})}{q(\Lambda|\boldsymbol{d})},
\end{equation}
where $\mathcal{L}(\Lambda|\boldsymbol{d})$ represents the classical hyper-likelihood given in \eqref{eq:tot_likelihood}. The reweighted posterior, shown in orange (DL with IS), aligns well with the classical result (HBA). While this process increases the computational cost of our method, it still requires significantly fewer likelihood evaluations than the standard HBA approach.
Additionally, importance sampling serves as a validation step: if the posterior remains unchanged after reweighting, it indicates that the model has successfully learned the correct HBA distribution.




Lastly, we comment that although the product of a normalizing flow model is a direct approximation of the posterior, the evidence can be estimated as well through importance sampling, which is nothing but a Monte Carlo estimate \cite{DeSanti:2024oap}. More precisely
\begin{equation}
    \mathcal{Z}_\Lambda=\int{\rm d}\Lambda~\mathcal{L}(\boldsymbol{d}|\Lambda)
    \pi(\Lambda)=\int{\rm d}\Lambda\frac{\mathcal{L}(\boldsymbol{d}|\Lambda)
    \pi(\Lambda)}{q(\Lambda|\boldsymbol{d})}
    q(\Lambda|\boldsymbol{d}),
\end{equation}
By sampling the flow posterior $q(\Lambda|\boldsymbol{d})$, which is optimized by minimizing the mass covering
forward KL divergence, we can get an estimator of the evidence from importance sampling
weights:
\begin{equation} \label{eq:evidence_is}
    \mathcal{Z}_\Lambda\approx\frac{1}{N}\sum_{i=1}^N\frac{\mathcal{L}(\boldsymbol{d}|\Lambda_i)
    \pi(\Lambda_i)}{q(\Lambda_i|\boldsymbol{d})}.
\end{equation}
The only disadvantage is that \eqref{eq:evidence_is} relies on the
analytical likelihood to be computed. As metioned before, it still requires significantly fewer likelihood evaluations than the conventional approach.
Besides, since they can be computed separately, the whole procedure can be
parallelized in principle, reducing its computational cost.


\section{Conclusions}
\label{sec:conclusions}

It can be expected that with next-generation GW detectors the
volume of data would rapidly grow, thus deep learning techniques
like the one presented here will become crucial for efficiently
extracting cosmological information from these observations.
Indeed, deep learning is already having a profound impact on GW
data analysis, spanning applications such as GW waveform modeling
\cite{McGinn:2021jqg,Liao:2021vec,Khan:2021czv,Islam:2022laz,Huerta:2017kez},
signal detection
\cite{2022PhRvD.105h3013M,Gabbard:2017lja,Wang:2019zaj,Ruan:2021fxq,Xia:2020vem,Krastev:2019koe},
parameter estimation
\cite{Green:2020hst,Shen:2019vep,Green:2020dnx,Krastev:2020skk,Sasaoka:2022hpt},
and so on. In this work, we have developed a novel deep learning
method to efficiently infer the population properties (of PBHs)
from GW population observations. Our method can significantly
reduce computational costs compared to conventional hierarchical
Bayesian analysis while maintaining comparable accuracy.

One of the key advantages of our approach is its ability to bypass
the need for explicit likelihood function construction, which is a
major computational bottleneck in conventional MCMC-based HBA
methods. The ability to rapidly estimate the PBH fraction in dark
matter $f_{\rm PBH}$ and other key population parameters allows
for real-time population analysis, which could be crucial for
distinguishing between astrophysical and primordial origins of
black hole mergers, making it particularly well-suited for
analyzing large GW datasets expected from space-based detectors,
such as LISA \cite{2017arXiv170200786A}, Taiji, \cite{Hu:2017mde},
Tianqin \cite{TianQin:2015yph}, and the next-generation
ground-based detectors. This computational efficiency is crucial
as the number of GW detections continues to grow, enabling
real-time population analysis and rapid updates to cosmological
constraints. Moreover, the flexibility of our method indicate that
it can be extended to incorporate more complex models, including
mixed PBH-ABH populations, accretion effects, and spin
distributions. This adaptability is particularly important as the
next-generation GW detectors may reveal more nuances of black hole
populations, requiring more sophisticated models to capture their
behavior.

There is still much room for improvement. The accuracy of our
method depends on the quality and coverage of the training
dataset, and systematic biases may arise if real GW events fall
outside the parameter space explored during training. It is also
noteworthy that while normalizing flows provide a powerful density
estimation technique, further improvements in architecture design
and regularization strategies could enhance robustness. How to
extend our method to multi-messenger observations, incorporating
real GW data, and further refining the underlying the models of
PBHs is also an issue worth focusing on. In particular, the
inclusion of spin distributions and more detailed accretion models
could provide a more comprehensive understanding of PBH
populations.


\section*{Acknowledgments}

This work is supported by National Key Research and Development
Program of China, No. 2021YFC2203004, and NSFC, No.12075246. We
acknowledge the use of high performance computing services
provided by the International Centre for Theoretical Physics
Asia-Pacific cluster, the Tianhe-2 supercomputer and Scientific
Computing Center of University of Chinese Academy of Sciences. We
would also like to thank the Dingo Collaboration team
\url{https://github.com/dingo-gw/dingo}.


\bibliography{REF}

\end{document}